\documentclass[12pt]{article}
\usepackage{graphicx}
\begin{document} 

\centerline{\bf Language simulation after a conquest}

\bigskip
Christian Schulze and Dietrich Stauffer\\
Institute for Theoretical Physics, Cologne University\\D-50923 K\"oln, Euroland

\bigskip
Abstract: When a region is conquered by people speaking another language, we 
assume within the Schulze model that at each iteration each person with 
probability $s$ shifts to the conquering language. The time needed for
the conquering language to become dominating is about $2/s$
for directed Barab\'asi-Albert networks, but diverges on the square lattice for 
decreasing $s$ at some critical value $s_c$.

\bigskip

The language competition model of Abrams and Strogatz assumes the possibility 
that of two competing languages one has a higher status \cite{Abrams}. We now 
look for an analogous question in the multi-language Schulze model 
\cite{langssw}: Will the language of the conquerors finally always win?

We used the Schulze model with $F$ features, each of which takes an integer value
between 1 and $Q$. All speakers are positioned on a directed Barab\'asi-Albert 
network of $N$ people surrounding a fully connected core of $m=3$ nodes. Each 
node added to the network selects $m$ already existing nodes as teachers, via
preferential attachment. At
each iteration, with probability $p = 0.5$ each feature of each node is 
modified. For this modification, with probability $q = 0.85$ the feature value
of a randomly selected neighbour (teacher) is taken, while with probability $1-q$ a 
randomly selected new value between 1 and $Q$ is taken. Also, at each iteration
each speaker with probability $(1-x)^2r \quad (r=0.9)$ gives up the old language
and takes over the language of a randomly selected neighbour; here $x$ is the 
fraction of the whole population speaking the old language.

Initially the $N$ people surrounding the core select their own language 
randomly, while the $m$ core members select the ``central'' language where
each feature is 2 for $Q=3$ and 3 for $Q=5$. Later, the core members are not 
subject to the above modifications and represent a ``royal family'' speaking
the unmodified official language. As a result, after a few iterations nearly
everybody speaks the central language of the core.

The influence of war is simulated as follows: A foreign power, speaking a 
language where all features are 1, conquers the country during ten iterations. 
From then on everybody, including the core, at each iteration with 
probability $s$ adopts the language of the conquerer. The effect of this 
language shift is particularly drastic on the whole population if a core 
member shifts to the conquering language. The winning time is defined as 
the total number of iterations (including the initial ten iterations of war) 
needed for the conquering language to become the numerically strongest 
language in the population and for {\it all} core members to adopted it. 

\begin{figure}[hbt]
\begin{center}
\includegraphics[angle=-90,scale=0.5]{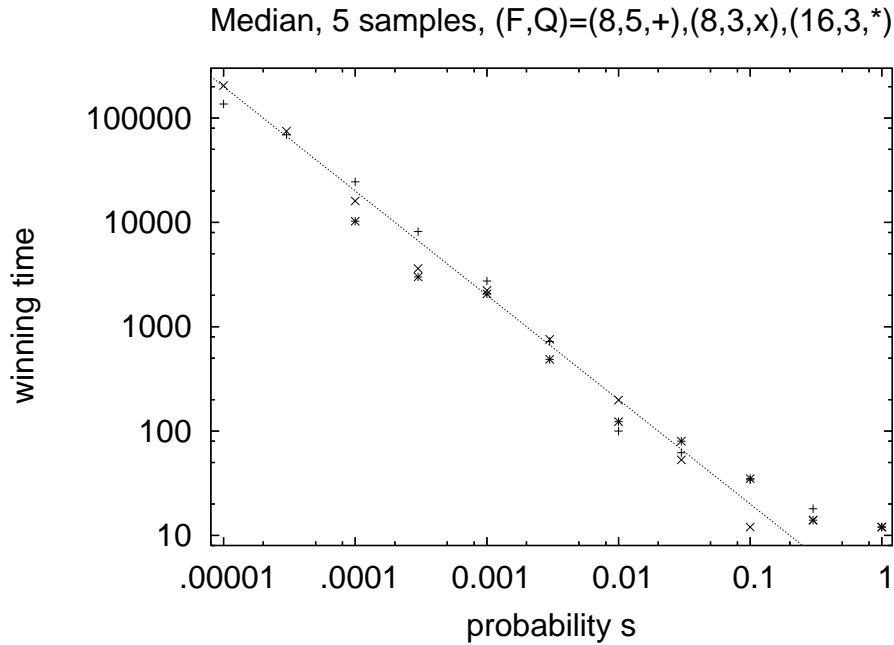}
\end{center}
\caption{Winning time versus adoption probability $s$, for $N = 10^4$.
The deviations for small times come from the initial time of war, 10 iterations,
after which the process of adopting the conquering language begins.
The straight line gives $2/s$.
}
\end{figure}

\begin{figure}[hbt]
\begin{center}
\includegraphics[angle=-90,scale=0.32]{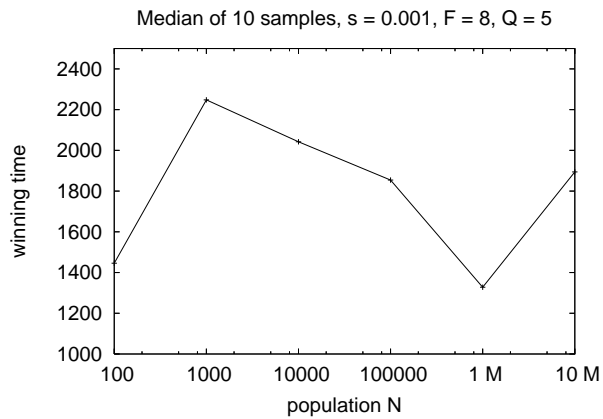}
\end{center}
\caption{Independence of the winning time of the population size $N$, at
$s = 10^{-3}, \; F = 8, \; Q = 5$.
}
\end{figure}

\begin{figure}[hbt]
\begin{center}
\includegraphics[angle=-90,scale=0.36]{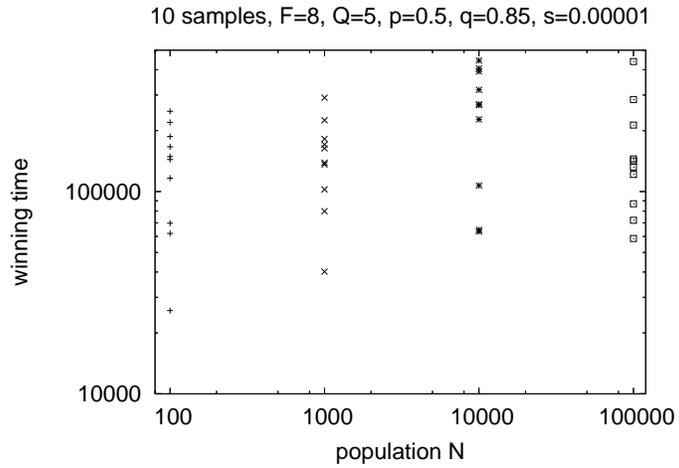}
\end{center}
\caption{Distribution of winning times at $s = 10^{-5}, \; F = 8, \; Q = 5$
versus population size $N$.
}
\end{figure}

\begin{figure}[hbt]
\begin{center}
\includegraphics[angle=-90,scale=0.36]{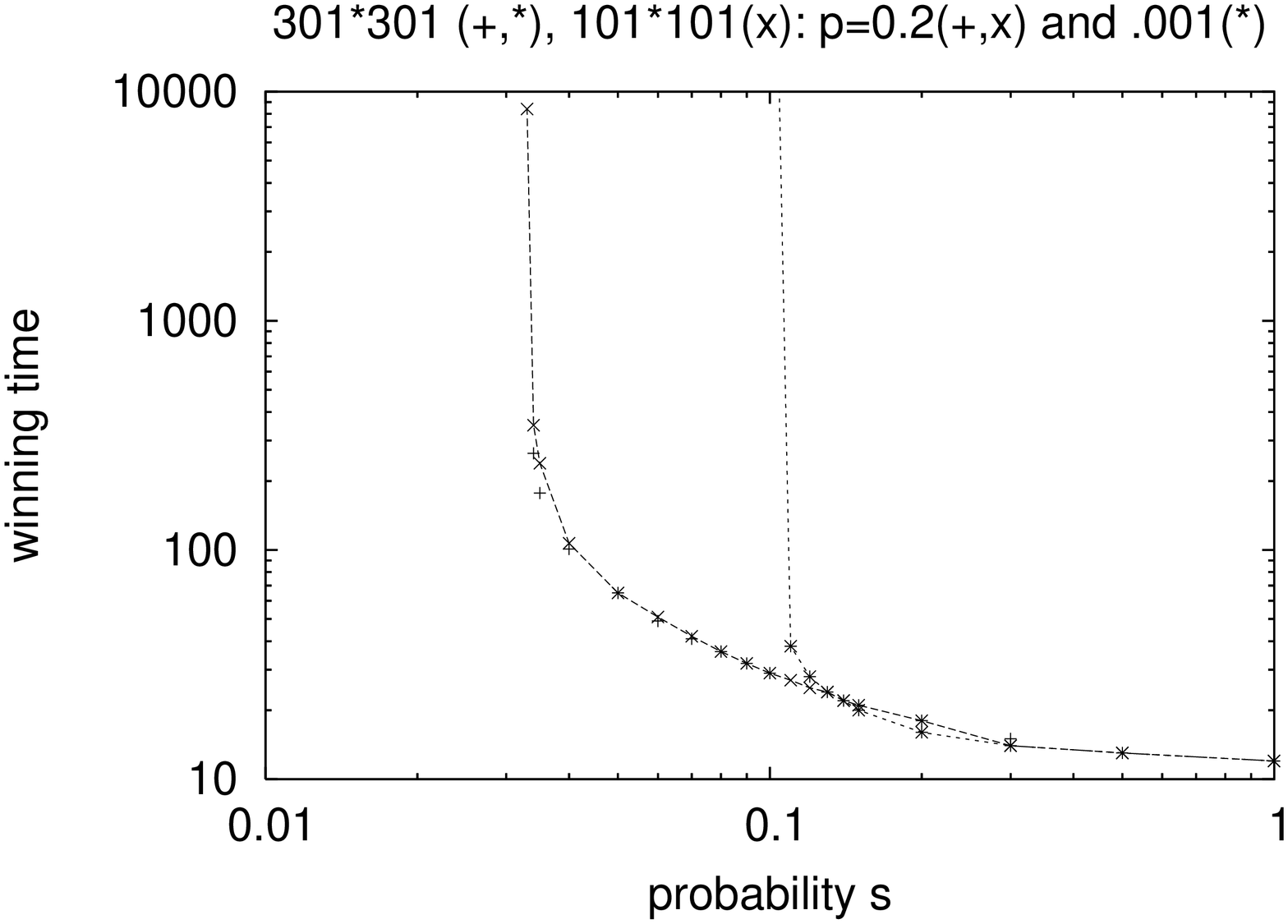}
\end{center}
\caption{Winning time versus adoption probability $s$ on square lattices,
showing a divergence independent of lattice size. (Median from 5 samples.)
}
\end{figure}

Fig.1 shows for three choices of $F$ and $Q$ that the winning time is
about inversely proportional to the probability $s$. It seems to be independent
of $N$, Fig.2. The fluctuations in the winning time are very large and do 
not seem to diminish if $N$ increases, Fig.3. Varying $p$ at fixed $q=0.85$, 
or $q$ at fixed $p=0.5$ does not change much (not shown).

In summary, there seems to be no threshold for $s$ in the winning times.
Even for very small $s$ the conquering language will win, after a time of
the order of $2/s$. Reality is, of course, more complicated that this model.
The Basque language is still used in Northeastern Spain thousands of years
after the neighbours started to speak an Indo-European language, while Celtic 
France mostly started to speak Latin and French only very few centuries after 
the Roman conquest. Within this model these differences would require that
different populations have different probabilities $s$ to adopt the conquering
language.

A rather different picture is obtained if we put the speakers on a square
lattice instead of the Barab\'asi-Albert network. Then Fig.4 shows rather small
time with little sample-to-sample fluctuations, diverging at some critical
value for the probability $s$. (Numerically, divergence means a median time
above one million.) Thus perhaps the Basque country in this version had an $s$ 
below this critical value ever since the Indo-European settlement of the
Iberian peninsula, while the $s$ for France was higher.

We thank L. Litov for suggesting this work and S. Wichmann for comments.

\end{document}